\documentclass[10pt]{wlscirep}

\usepackage{multirow}
\usepackage{array}
\usepackage{threeparttable}

\title{High speed error correction for continuous-variable quantum key distribution with multi-edge type LDPC code}

\author[1]{Xiangyu Wang}
\author[1,$\dag$]{Yichen Zhang}
\author[1,*]{Song Yu}
\author[2]{Hong Guo}
\affil[1]{State Key Laboratory of Information Photonics and Optical Communications, Beijing University of Posts and Telecommunications, Beijing, 100876, China}
\affil[2]{State Key Laboratory of Advanced Optical Communication Systems and Networks, School of Electronics Engineering and Computer Science, Center for Quantum Information Technology, Center for Computational Science and Engineering, Peking University, Beijing, 100871, China}
\affil[$\dag$]{zhangyc@bupt.edu.cn}
\affil[*]{yusong@bupt.edu.cn}

\begin{abstract}
Error correction is a significant step in postprocessing of continuous-variable quantum key distribution system, which is used to make two distant legitimate parties share identical corrected keys. We propose an experiment demonstration of high speed error correction with multi-edge type low-density parity check (MET-LDPC) codes based on graphic processing unit (GPU). GPU supports to calculate the messages of MET-LDPC codes simultaneously and decode multiple codewords in parallel. We optimize the memory structure of parity check matrix and the belief propagation decoding algorithm to reduce computational complexity. Our results show that GPU-based decoding algorithm greatly improves the error correction speed. For the three typical code rate, i.e., 0.1, 0.05 and 0.02, when the block length is $10^6$ and the iteration number are 100, 150 and 200, the average error correction speed can be respectively achieved to 30.39Mbits/s (over three times faster than previous demonstrations), 21.23Mbits/s and 16.41Mbits/s with 64 codewords decoding in parallel, which supports high-speed real-time continuous-variable quantum key distribution system.
\end{abstract}

\begin{document}

\flushbottom
\maketitle

\thispagestyle{empty}

\section*{Introduction}

Quantum key distribution (QKD)~\cite{QKD} allows two legitimate parties Alice and Bob to share unconditional security keys through an untrusted quantum channel and a classical authenticated channel, even if in the presence of an eavesdropper Eve. Many QKD protocols have been proposed since the first QKD protocol was proposed in 1984, they encode the key information on discrete variables (DV)~\cite{BB84,E91} (such as the polarization or phase of single photon pulses) or continuous variables (CV)~\cite{GG02,2004NOSW} (such as the quadratures of coherent states). Compared to DV protocols, CV protocols use homodyne detector or heterodyne detector to measure the quantum states, which have the advantage of using standard telecommunication technologies~\cite{2012RMP,2015en}. Recently, a new CV protocol design framework (LZG framework) has been proposed to allow one to design the protocol using arbitrary non-orthogonal states with their application scenarios~\cite{2018LI}. CV-QKD protocols eliminate the limitation of single photon detector and have more advantages in practical QKD protocols.

For a practical Gaussian-modulated coherent state CV-QKD system, the speeds of information reconciliation and privacy amplification have an important influence on the secret key rate, and the efficiency of information reconciliation affects the secret key rate and transmission distance~\cite{Paul13,Zhang2017,long11}. High speed and high performance reconciliation has been studied in DV-QKD~\cite{MM2013,AR2014}. High speed privacy amplification has also been implemented~\cite{WangPA}. However, the speed of information reconciliation still limits the performance of CV-QKD systems. Due to the raw keys of Alice and Bob are correlated Gaussian variables, some approaches~\cite{Slice1,MD} have been proposed to achieve excellent efficiency. Multidimensional reconciliation~\cite{MD} obtains high efficiency at low signal-to-noise ratios (SNR) by rotating the Gaussian variables to construct an virtual binary input additive white Gaussian noise channel. The error correction performance of multi-edge type low-density parity check codes (MET-LDPC)~\cite{METLDPC,CODE} are close to Shannon limit. Multidimensional reconciliation and MET-LDPC codes can be combined to achieve excellent efficiency at low SNRs~\cite{Paul13,long11,WANGRA}, which supports  CV-QKD system. Thus, we mainly focus on accelerating the speed of information reconciliation.

As previously described, information reconciliation contains two processes: multidimensional reconciliation and error correction with MET-LDPC codes. The computational complexity of the first process is low, which can achieve high speed on central processing unit (CPU). However, for decoding with CPU, the speed of the error correction process will be quite slow when MET-LDPC codes approach to the Shannon limit at low SNRs~\cite{Polar}. The main reasons are that: 1) the computational complexity of belief propagation decoding algorithm is high for long-block-length (on the order of $10^6$) and low-code-rate (no higher than 0.1) MET-LDPC codes; 2) belief propagation decoding algorithm requires more iterations to converge at low SNRs. Several work have been proposed to speed up the error correction process. They achieve the decoding speed to 7.1Mb/s~\cite{Polar} and 9.17Mb/s~\cite{Mili917} with LDPC codes based on graphic processing unit (GPU). 

In this paper, we propose a high speed parallel multiple codewords MET-LDPC code error correction method based on GPU. We optimize the memory structure of parity check matrix, making the decoding process more efficient. We modify the belief propagation decoding algorithm, which reduces computational complexity. This work has been applied to the longest field test of a CV-QKD system and achieves secure key rates two orders-of-magnitude higher than previous field test demonstrations~\cite{Zhang2017}.

\section*{Results}

\textbf{Information reconciliation for CV-QKD system.}  Information reconciliation is an efficient way for Alice and Bob to distill common corrected keys from their related variables. In a Gaussian-modulated coherent state CV-QKD system, the raw keys of Alice and Bob are continuous variables which cannot directly use the channel coding technology to correct errors between them. To solve this problem, several work have been done to extract common string from Gaussian variables. Sign reconciliation~\cite{SIGN} encodes information on the sign of Gaussian variables. However, since most Gaussian values are close to 0 at low SNRs, it is difficult to distinguish the sign of Gaussian variables. In Ref.~\cite{SIGN}, they only use high-amplitude data by post-selection, but this method discards a large number of small-amplitude data, which reduces the data utilization rate. Another method called slice reconciliation~\cite{Slice1,Slice2} divides Gaussian variables to different slices and then encodes information on the quantized slices. Due to the limitation of efficiency, this method is applicable to short distance CV-QKD system. In Ref.\cite{MD}, they rotate the Gaussian variables to construct a virtual binary input additive white Gaussian noise channel, then Alice and Bob's Gaussian variables will be converted to a binary string and the noise form of this binary string respectively. This method is called multidimensional reconciliation which is suitable for CV-QKD system.

Information reconciliation has two modes: direct reconciliation and reverse reconciliation~\cite{RR}. Due to the limitation of 3dB loss, the maximum transmission distance of direct reconciliation algorithm is 15km when the optical fiber loss is 0.2dB/km. However, reverse reconciliation algorithm can break this limit. In order to achieve long distance and high secret key rate of CV-QKD system, efficient error correction codes are required to distill secret keys from Alice and Bob's correlated Gaussian variables at low SNRs. MET-LDPC codes~\cite{CODE} are one of the error correction codes, which have well error correction performance even if at low SNRs.

For CV-QKD system, we combine multidimensional reconciliation and MET-LDPC codes to obtain excellent reconciliation efficiency at low SNRs by using reverse reconciliation protocols. Assuming that the Gaussian variables of Alice follow a zero mean and $\sigma_{X}^{2}$ variance Gaussian distribution $X \sim \mathcal{N}(0,\sigma_{X}^{2})$, Bob's Gaussian variables $Y \sim \mathcal{N}(0,\sigma_{Y}^{2})$ and the quantum channel noise $Z \sim \mathcal{N}(0,\sigma_{Z}^{2})$, where $\sigma_{Y}^{2}=\sigma_{X}^{2}+\sigma_{Z}^{2}$, and $Y=X+Z$. In order to achieve effective error correction at low SNRs, Bob and Alice first use multidimensional reconciliation to convert their Gaussian variables $Y$ and $X$ to binary string $U$ and the noise form $V$ of this binary string. Then Alice and Bob correct their errors with MET-LDPC codes based on belief propagation decoding algorithm. Finally, they share a common binary string $U$ with a certain probability. The secret key rate of CV-QKD system can be calculated by $k=\beta I(x;y)-S(y;E)$, where $\beta$ is the efficiency of information reconciliation, $I(x;y)$ is the Shannon entropy of Alice and Bob, $S(y;E)$ is the Von Neumann entropy of Bob and Eve.

\begin{figure}[t]
\centering\includegraphics[width=13.3cm]{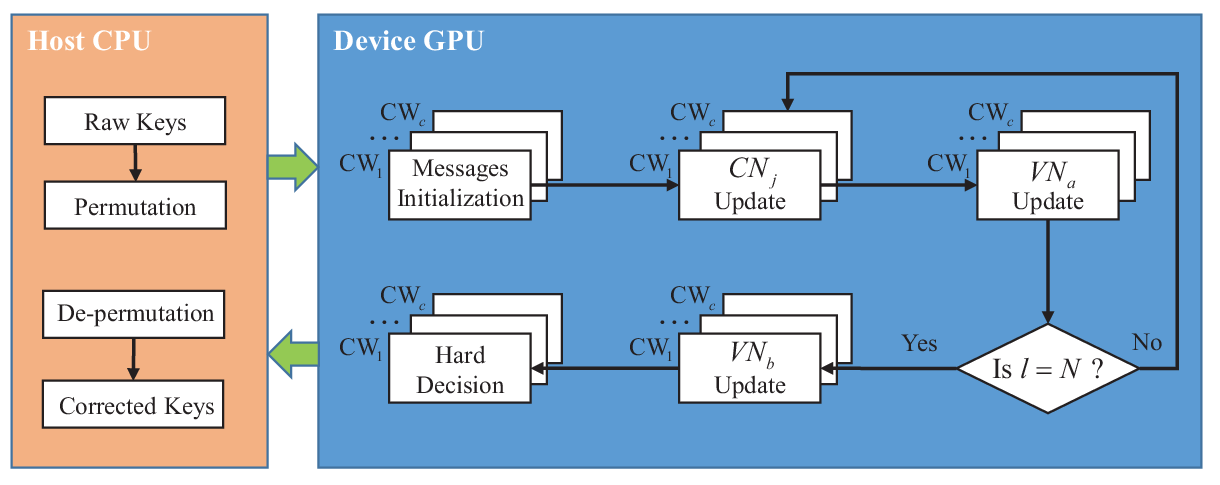}
\caption{The process of GPU-based multiple codewords parallel decoding algorithm. CW$_{1}, \cdots,$ CW$_{c}$ represent different codewords. c stands for the number of codewords decoded in parallel. CN$_{j}$ represents the set of check nodes. VN$_{a}$ and VN$_{b}$ represent the sets of the variable nodes whose degrees are greater than 1 and equal to 1. $l$ and $N$ stand for the current number of iterations and maximum number of iterations, respectively.}
\label{GPU-based decoder}
\end{figure}

\textbf{High speed error correction with MET-LDPC codes.}  High speed error correction is required to support real-time CV-QKD system. The error correction speed of MET-LDPC codes is related to the decoding algorithm, code length, the number of iterations, implementation method and other factors. For  CV-QKD system, the error correction is quite difficult due to the low SNRs. Thus, we have to choose belief propagation decoding algorithm which iteratively updates message between variable nodes and check nodes to converge on valid codewords. The code length of a codeword is on the order of $10^6$. When the MET-LDPC codes near to the Shannon limit, the reconciliation efficiency approaches to 1, the decoder needs more iterations to converge. In order to achieve high speed error correction at low SNRs, we implement the MET-LDPC decoder on GPU platform which supports to update the messages of variable nodes and check nodes in parallel. To maximize the parallel performance of GPU, we propose a method for simultaneously decoding multiple codewords. We also modify the belief propagation decoding algorithm and optimize the memory structure of parity check matrix to further accelerate the error correction process.

The decoding speed is extremely slow for long code length at low SNRs when we perform the decoder on CPU. Thus, we implement the MET-LDPC multiple codewords decoder on GPU with compute unified device architecture application programming interface developed by NVIDIA corporation~\cite{CUDA}. The GPU-based parallel decoding process is shown in Figure~\ref{GPU-based decoder}. We first copy the messages of permutated raw keys from host (CPU) to device (GPU). Then we initialize the messages of variable nodes and check nodes with kernel function on GPU. Next, we build two kernel functions to iteratively update messages of check nodes and variable nodes. It is not necessary to update all the variable nodes, the iteration process only update probabilities messages of all check nodes and the variable nodes whose degree is greater than 1 without making hard decisions. In our GPU-based decoding process, we ignore the variable nodes whose degree is equal to 1, this will reduce computational complexity and save a large number of threads. Without hard decision, the decoder will be simplified. After the maximum number of iterations is reached, the LDPC decoder calculates the probability messages of the variable nodes whose degree is equal to 1 and then performs hard decisions to get the decoded data and copy them from device to host. Finally, we de-permutate the decoded data to obtain the corrected keys.

\begin{figure}[t]
\centering\includegraphics[width=13.35cm]{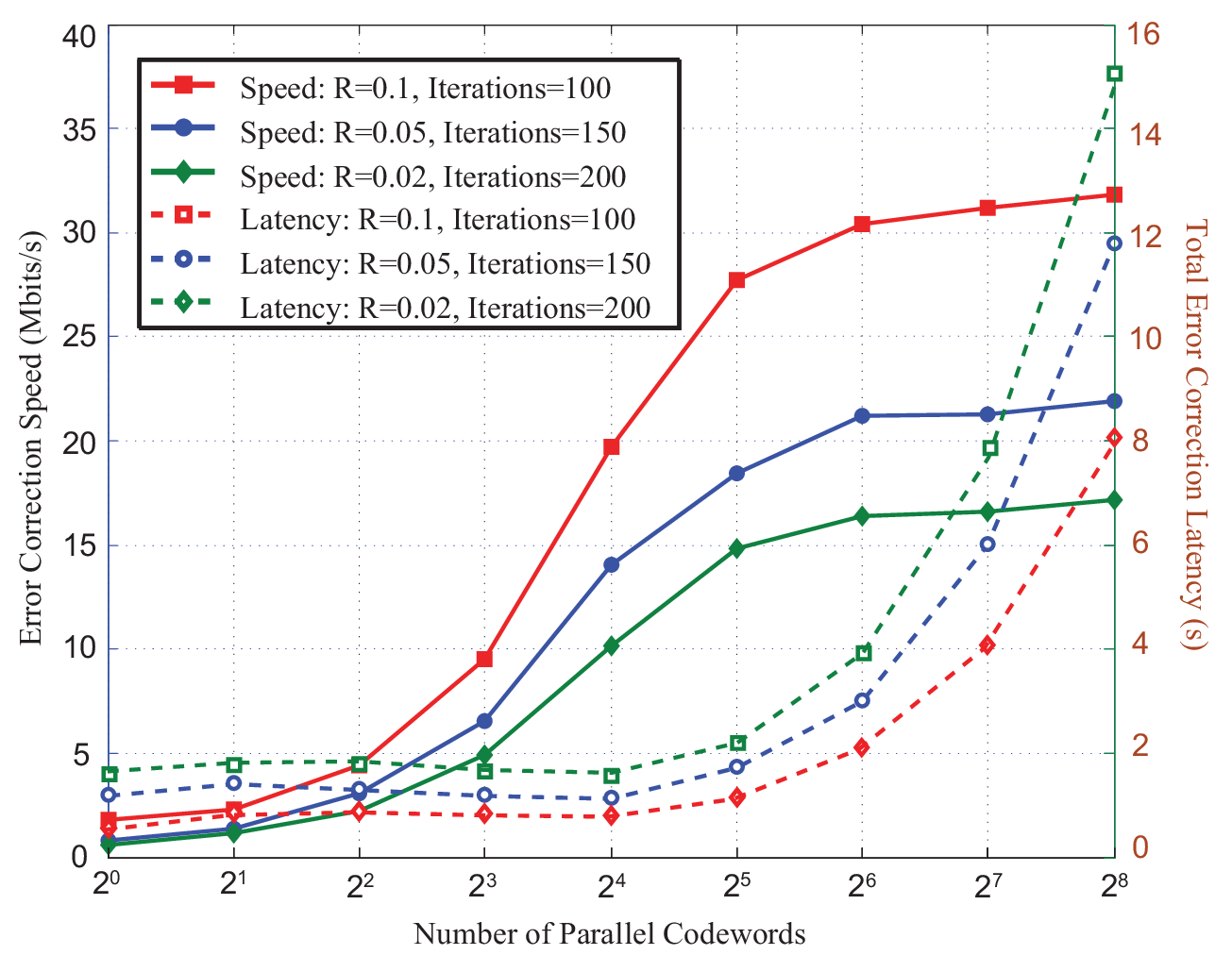}
\caption{The error correction speeds and latency of different number of codewords to decode in parallel with rate 0.1, 0.05 and 0.02 code and the iterations are 100, 150, and 200 respectively. The solid lines refer to the error correction speed. The dotted lines refer to the total error correction latency. The block length is $10^6$. The results are obtained on a NVIDIA TITAN Xp GPU.}
\label{threeresults}
\end{figure}

For NVIDIA TITAN Xp GPU, the maximum number of thread blocks and threads per thread block on a grid are 65536 and 1024 respectively. Thus the maximum data that can be simultaneously decoded is 65536*1024=67108864. However, the block length of a codeword is $10^6$ in our system. The parallel performance of the GPU can not be fully exploited when decoding with only one codeword. We can further accelerate the error correction speed by parallel decoding with multiple codewords. According to the parameters of GPU, we calculate that 64 codewords can be simultaneously decoded at most. Actually, since only the messages of the variable nodes whose degree is greater than 1 are updated, there are still a large number of threads that can be allocated when the messages of variable nodes are updated. For updating the check nodes messages, we reuse the threads that have been performed. Therefore, the number of simultaneous decoding codewords can be greater than 64. Theoretically, any number of codewords is possible as long as the GPU has enough memory.

The latency of global memory access has a significant impact on error correction speed. Coalesced global memory access can hide the latency. However, in order to obtain excellent reconciliation efficiency, the parity check matrix $H$ is randomly constructed, where $H$ is a two-dimensional matrix. And the block length of $H$ is very long, we have to allocate the messages in global memory. When the decoder updates messages, the latency of non-consecutive global memory access of $H$ limits the MET-LDPC error correction speed. No matter whether updating the variable nodes or the check nodes messages, the read and write access to global memory is non-consecutive because that both the variable nodes and check nodes of $H$ are unorder. The latency can be hidden by optimizing the memory structure of $H$. We store $H$ in two files, one of which stores variable nodes sequentially, and the other stores the mapping relations of variable nodes to check nodes. We can also swap variable nodes and check nodes. In this way, memory access for variable nodes will be consecutive. For simultaneous decoding of multiple codewords, the raw key permutation enables the memory access of check nodes to become consecutive. The kernels of GPU are performed by warps. A wrap contains multiple threads which perform the same program instruction in parallel, but with different data. Different type of GPU has different number of threads in a wrap. Typically, it is 32. If the threads inside a wrap access consecutive global memory, the latency will be hidden. Thus, when the number of simultaneously decoded codewords is an integer multiple of 32, both of the variable nodes and check nodes memory access are consecutive. Actually, when the latency of memory access equal to the latency of the messages update, the error correction speed is no longer improved by increasing the number of parallel codewords. By simultaneously decoding with multiple codewords based on GPU, the error correction speed is greatly improved, which supports high speed real-time CV-QKD system.

\begin{table}[t]
\centering
\begin{threeparttable}
\centering
\renewcommand\arraystretch{1.21}
\small
\caption{GPU-based error correction speed and error correction performance with 64 codewords parallel decoding. SNR: signal-to-noise ratio. $\beta$: reconciliation efficiency.}
\label{ResultsOfThreeCodes}
\begin{tabular}{|*{7}{c|}}
\hline
Code Rate & \multicolumn{2}{c|}{0.1} & \multicolumn{2}{c|}{0.05} & \multicolumn{2}{c|}{0.02} \\ \hline
SNR & \multicolumn{2}{c|}{0.160} & \multicolumn{2}{c|}{0.075} & \multicolumn{2}{c|}{0.029} \\ \hline
$\beta$ & \multicolumn{2}{c|}{93.40\%} & \multicolumn{2}{c|}{95.84\%} & \multicolumn{2}{c|}{96.99\%} \\ \hline
Iterations & \multicolumn{2}{c|}{100} & \multicolumn{2}{c|}{150} & \multicolumn{2}{c|}{200} \\ \hline
FER & \multicolumn{2}{c|}{0.055} & \multicolumn{2}{c|}{0.203} & \multicolumn{2}{c|}{0.375} \\ \hline
Total Number & \multicolumn{2}{c|}{ \multirow{2}{*}{3,767,500}} & \multicolumn{2}{c|}{ \multirow{2}{*}{3,480,000}} & \multicolumn{2}{c|}{ \multirow{2}{*}{3,337,500}} \\
of Edges & \multicolumn{2}{c|}{} & \multicolumn{2}{c|}{} & \multicolumn{2}{c|}{} \\ \hline
Updated CNs & \multicolumn{2}{c|}{900,000} & \multicolumn{2}{c|}{950,000} & \multicolumn{2}{c|}{980,000} \\ \hline
Updated VNs & 1,000,000 & 125,000 & 1,000,000 & 70,000 & 1,000,000 & 40,000 \\ \hline
Ignored VNs$^1$ & 0 & 875,000 & 0 & 930,000 & 0 & 960,000 \\ \hline
Number of Edges &  \multirow{3}{*}{3,767,500} & \multirow{3}{*}{2,892,500} & \multirow{3}{*}{3,480,000} & \multirow{3}{*}{2,550,000} & \multirow{3}{*}{3,337,500} & \multirow{3}{*}{2,377,500} \\
to pass messages &&&&&& \\
(CNs to VNs)$^2$ &&&&&& \\ \hline
Latency Per &  \multirow{2}{*}{0.363} & \multirow{2}{*}{0.329} & \multirow{2}{*}{0.361} & \multirow{2}{*}{0.314} & \multirow{2}{*}{0.357} & \multirow{2}{*}{0.305} \\
Iteration (ms)$^3$ &&&&&& \\ \hline
Error Correction &  \multirow{2}{*}{27.54} & \multirow{2}{*}{30.39} & \multirow{2}{*}{18.49} & \multirow{2}{*}{21.23} & \multirow{2}{*}{14.00} & \multirow{2}{*}{16.41} \\
Speed (Mbits/s)$^4$ &&&&&& \\
\hline
\end{tabular}
\begin{tablenotes}
\item[1] These VNs (variable nodes) are ignored only when the decoder performs the iterative process. Their messages will be computed before the hard decision process.
\item[2] Because only the VNs have degree 1, the number of edges to pass messages would be reduced only when the messages pass from CNs (check nodes) to VNs.
\item[3] The latency per iteration is an average for total decoding latency, including the latency of initialization, iterative message-passing, CNs and VNs updated, hard decision and memory copy between CPU and GPU.
\item[4] The results are obtained on a NVIDIA TITAN Xp GPU.
\end{tablenotes}
\end{threeparttable}
\end{table}

\textbf{GPU-based error correction speed.}  We implement high speed error correction with multiple codewords based on GPU. For  CV-QKD system, we choose low-code-rate MET-LDPC codes to correct error at low SNRs. Three typical code rates are designed in this work, {\it i.e.}, 0.1, 0.05 and 0.02, we all achieve high error correction speed on long block length and high iteration number. The block length of each code is $10^6$. For different codes, the number of iteration are uncertain because that they apply to different distances (Actually, it is mainly affected by SNRs). We apply these three codes to correct errors when the SNR are 0.161, 0.075 and 0.029 and we set the iteration number to 100, 150 and 200, respectively. The degree distribution of these three codes are proposed in \cite{WANGRA}. The parity check matrices are randomly constructed by progressive edge growth algorithm. In Figure~\ref{threeresults}, we show the error correction speed of different number of codewords simultaneous decoding.

As shown in Figure~\ref{threeresults}, when the number of codewords is less than 32, the error correction speeds increase rapidly. The main reason is that the latency is hidden by coalesced global memory access. When the threads in a wrap access non-consecutive global memory, the latency will be very long, even longer than updating the messages. Thus, the GPU-based decoder spends almost the same time when the number of codewords is less than 32. In other words, the total time is almost the same, either waiting for memory access or updating the messages. The error correction speed will be no longer improved by increasing the number of codewords when the access memory latency is the same as updating messages latency. Only by simplifying the decoding computational complexity can we further accelerate the error correction speed. As shown in Figure~\ref{threeresults}, the error correction speed is almost no longer increased when the number of codewords is greater than 64. The requirement for CPU and GPU are too much if the number of codewords is too large. After comprehensive consideration, we choose 64 codewords to decode in parallel.

Table~\ref{ResultsOfThreeCodes} gives the GPU-based error correction speed and error correction performance of the three codes at low SNRs. For the rate 0.1, 0.05 and 0.02 codes with block length $10^6$, we achieve the error correction speeds to 30.39Mbits/s, 21.23Mbits/s and 16.41Mbits/s when the maximum number of iterations are 100, 150 and 200, respectively. The corresponding SNRs are 0.160, 0.075, and 0.029, the reconciliation efficiencies can be achieved to 93.4\%, 95.84\%, and 96.99\% respectively. The frame error rate (FER) indicates the error correction performance of MET-LDPC codes, it refers to the failure probability of error correction. For the implementation of the three code rates, they are 0.055, 0.203, and 0.375. Moreover, the failure probabilities can be reduced by increasing the maximum number of iterations.

\section*{Discussion}

We propose an experiment implement of GPU-based high speed error correction for CV-QKD system. A Multiple codewords parallel belief propagation decoder is presented to accelerate the iterative message-passing algorithm. For belief propagation decoding algorithm, the computational complexity of MET-LDPC codes originates from the number of connected edges between variable nodes and check nodes and the number of iterations. High error correction performance is required for CV-QKD system, we can not reduce the complexity by simplifying the decoding algorithm or shortening the block length. To reduce the computational complexity, we optimize the decoder by ignoring the variable nodes whose degree is equal to 1 when the decoder iteratively passes messages. These nodes do not affect message-passing. The messages of these variable nodes are computed after the iterative process. To hide the latency of the decoder, we modify the memory structure of parity check matrix so that the global memory access becomes consecutive.

\begin{table}[t]
 \centering
 \caption{Error correction speed comparison by different type of codes.}
 \newcommand{\tabincell}[2]{\begin{tabular}{@{}#1@{}}#2\end{tabular}}
 \renewcommand\arraystretch{1.21}
\begin{tabular}{|*{6}{c|}}
    \hline
    Refs. & Code Type & \tabincell{c}{Block\\ Length} & \tabincell{c}{Average\\ Iterations} & \tabincell{c}{Latency Per\\ Iteration (ms)} & \tabincell{c}{Decoding Speed\\ (Mbits/s)} \\
    \hline
    Paul {\it et al.}\cite{Polar} & Polar & $2^{27}$ & 1 & 18.386 & 7.3  \\
    Paul {\it et al.}\cite{Polar} & MET-LDPC & $2^{20}$ & 100 & 1.477 & 7.1  \\
    Milicevic {\it et al.}\cite{Mili917} & QC-LDPC & $2^{20}$ & 78 & 1.466 & 9.17 \\
    This work & MET-LDPC & $10^6$ & 100 & 0.329 & 30.39  \\
    \hline
\end{tabular}
\label{ResultsComparison}
\end{table}

As shown in Table~\ref{ResultsComparison}, we compare the performance between the proposed GPU-based multiple codewords parallel decoding and the results obtained by other work with rate 0.1 code at SNR=0.161. Paul {\it et al.} respectively obtain the speed to 7.1Mbits/s with MET-LDPC code on GPU and 7.3Mbits/s with Polar code on CPU~\cite{Polar}. The generator matrix of Polar codes have regular recursion structure. And the decoder is implemented by successive cancellation algorithm, which does not require iteration. However, the Polar decoder can not be implemented on GPU because that the nodes are associated when using successive cancellation algorithm. Milicevic {\it et al.} obtain the speed to 9.17Mbits/s with quasi-cyclic (QC) LDPC codes and early termination of the iteration process~\cite{Mili917}. QC-LDPC codes simplify the randomness connection of parity check matrix. However, the error correction performance will be decreased when the expansion factor is too large. The early termination scheme is an efficient way to reduce the complexity of LDPC decoder and avoids unnecessary iterations. On the contrary, the complexity of decoder will be increased if we use the early termination scheme to multiple codewords parallel decoding because that the early termination condition of each codeword is different. The error correction speed we achieved is over three times faster than previous demonstrations, which is supporting high speed real-time continuous-variable quantum key distribution system~\cite{Zhang2017}.

\section*{Methods}

\textbf{Belief propagation decoding algorithm of LDPC code.}  LDPC coeds~\cite{CODE,LDPC62} are block error correction codes with a sparse parity check matrix proposed by Gallager in 1962. Its error correction performance is close to Shannon limit. MET-LDPC codes~\cite{CODE} are generalization form of LDPC codes, which has better error correction performance even if at low SNRs. Typically, LDPC code is defined by parity check matrix $H$ of size $m \times n, m<n$. The code rate is defined by $R=(n-m)/n$. LDPC codes also can be represented by bipartite factor graphs~\cite{Tanner}. For a parity check matrix, $m$ represents the number of check nodes and $n$ represents the number of variables nodes. The variable nodes and check nodes are connected by edges. We use progressive edge-growth method~\cite{PEG} to construct parity check matrix based on the degree distribution proposed in \cite{WANGRA}. The MET-LDPC code decoding algorithm we used is belief propagation which iteratively propagates message between variable nodes and check nodes to converge on valid codewords until the decoding termination condition is satisfied or reaching to the maximum number of iterations. The belief propagation decoding algorithm in the reverse reconciliation postprocessing of CV-QKD system is described as follows.

Let $R_{j}$ be the set of variable nodes that are connected to the $j$th check node, $C_{i}$ be the set of check nodes that are connected to the $i$th variable node, $R_{j}\backslash i$ be the set $R_{j}$ excludes $i$, $C_{i}\backslash j$ be the set $C_{i}$ excludes $j$, $q_{ij}$ be the message passed from $i$th variable node to $j$th check node, $r_{ji}$ be the message passed from $j$th check node to $i$th variable node.

{\it Step~1}: Bob calculates the syndromes $S_{B}$ of his binary string that is achieved by multidimensional reconciliation and sends the syndromes to Alice.

{\it Step~2}: Alice calculates the initialization probabilities $q_{ij}^{0}$ ($i=1,2,\cdot\cdot\cdot,n,~j=1,2,\cdot\cdot\cdot,m$) that binary input additive white Gaussian noise channel passes to variable nodes. The superscript represents the current number of iterations. Theoretically, since the information that we extract are binary strings, the initialization probabilities include the probability of 0 and 1. To simplify the computational complexity, we use the ratio of $q_{ij}^{0}(1)$ to $q_{ij}^{0}(0)$ to represents the initialization probability.
\begin{equation}
q_{ij}^{0}=\frac{q_{ij}^{0}(1)}{q_{ij}^{0}(0)}
\end{equation}

{\it Step~3}: Alice updates the messages of check nodes. For the $j$th check node and $R_{j}$, she calculates the messages that variable nodes pass to check nodes when the iteration number is $l,~l=1,2,\cdot\cdot\cdot,N$, where $N$ is the maximum number of iterations.
\begin{equation}
r_{ji}^{l}=\frac{r_{ji}^{l}(1)}{r_{ji}^{l}(0)}=\frac{1-t}{1+t}
\end{equation}
\begin{equation}
t=\prod_{i^{'}\in R_{j}\backslash i}\frac{1-q_{i^{'}j}^{l-1}}{1+q_{i^{'}j}^{l-1}}
\end{equation}

{\it Step~4}: Alice updates the messages of variable nodes. For the $i$th variable node and $C_{i}$, she calculates the messages that check nodes pass to variable nodes when the iteration number is $l$.
\begin{equation}
q_{ij}^{l}=\frac{q_{ij}^{l}(1)}{q_{ij}^{l}(0)}=q_{ij}^{0}\prod_{j^{'}\in C_{i}\backslash j}r_{j^{'}i}^{l}
\end{equation}

{\it Step~5}: Alice makes hard decisions. If $q_{i}^{l}>1$, the codeword $c_{i}=1$,  otherwise $c_{i}=0$. Alice calculates the syndrome $S_{A}$ of codeword $c$, such that $S_{A}=Hc^{T}$. If $S_{A}$ is equal to $S_{B}$ or reaching to the maximum number of iterations, the decoding is ended, otherwise repeat step 3 to step 5.
\begin{equation}
q_{i}^{l}=\frac{q_{i}^{l}(1)}{q_{i}^{l}(0)}=q_{ij}^{0}\prod_{j\in C_{i}}r_{ji}^{l}
\end{equation}

We can use log-likelihood ratios to represent the probabilities messages. This decoding algorithm converts a large number of multiplication into addition, which reduces the computational complexity of belief propagation algorithm. A lookup table can be built to accelerate the process of updating the messages of log-likelihood ratios.


\section*{Acknowledgements}

This work was supported by the Key Program of National Natural Science Foundation of China under Grant 61531003, the National Natural Science Foundation under Grant 61427813, the National Basic Research Program of China (973 Program) under Grant 2014CB340102, and the Fund of State Key Laboratory of Information Photonics and Optical Communications.

\section*{Author contributions}

H. G. and S. Y. proposed and guided the work. X. W. and Y. Z. designed and performed the experiment. All authors analysed the results and wrote the manuscript.

\section*{Additional information}

\textbf{Competing financial interests:} The authors declare that they have no competing interests.

\end{document}